\documentclass[lettersize,journal,twocolumn]{IEEEtran}

\usepackage[utf8]{inputenc}
\usepackage{amsmath}
\usepackage{amssymb}
\usepackage[english]{babel}
\usepackage[T1]{fontenc}
\usepackage{amsthm}
\usepackage{amsfonts}
\usepackage{graphicx}
\usepackage{tikz}
\usetikzlibrary{positioning,calc}

\usepackage[compatibility=false]{caption}
\usepackage{subcaption}
\usepackage{pgfplotstable}
\usepackage{nameref}
\usepackage{pgfplots}
\pgfplotsset{compat=1.18} 
\usepackage{filecontents}
\usepackage[table]{xcolor}
\usepackage{txfonts}
\usepackage{algorithm}
\usepackage[noend]{algpseudocode}

\usepackage{float} 
\usepackage{hyperref}

\usepackage{cancel}
\usepackage{graphicx}
\usepackage{array}
\usepackage{lipsum}
\usepackage{booktabs}
\usepackage{pythontex}
\usepackage{tabularx}
\usepackage{longtable}
\usepackage{hyphenat}
\usepackage{enumitem}
\usepackage{ulem}
\usepackage{setspace} 

\usepackage{hyperref}
\usepackage{cleveref}

\crefname{subfigure}{Figure}{Figures}
\Crefname{subfigure}{Figure}{Figures}
\crefname{figure}{Figure}{Figures}
\Crefname{figure}{Figure}{Figures}

\usepackage{booktabs}

\usepackage{tikz}

\usepackage{cite}

\usepackage[numbers,sort&compress]{natbib}

\usepackage{authblk}

\begin{document}

\author[1]{Sara Tarquini$^*$\thanks{*Email: sara.tarquini@gssi.it}}
\affil[1]{Gran Sasso Science Institute,  Viale Francesco Crispi 7, L'Aquila, Italy}
\author[2]{Matteo Vandelli}
\affil[2]{Quantum Computing Solutions, Leonardo S.p.A., Via R. Pieragostini 80, Genova, Italy}
\author[2]{Francesco Ferrari}
\author[2,3]{Daniele Dragoni}
\affil[3]{Hypercomputing Continuum Unit, Leonardo S.p.A., Via R. Pieragostini 80, Genova, Italy}
\author[1,4]{Francesco Tudisco}
\affil[4]{School of Mathematics and Maxwell Institute, University of Edinburgh, Peter Guthrie Tait Road, EH9 3FD, Edinburgh, UK}

\title{Drone delivery packing problem on a neutral-atom quantum computer}

\maketitle

\begin{abstract}
Quantum architectures based on neutral atoms have gained significant attention in recent years as specialized computational machines, due to their ability to directly encode the independent set constraint on graphs, exploiting the Rydberg blockade mechanism. In this work, we address the Drone Delivery Packing Problem via a hybrid quantum-classical framework leveraging a neutral-atom quantum processing unit (QPU). We reformulate the optimization task as a graph-partitioning problem based on the independent sets (ISs) of a scheduling graph that encodes delivery incompatibilities. Each partition corresponds to deliveries assigned to a single drone, with the objective of minimizing the total number of partitions. While the ISs represent time-feasible schedules, battery-duration constraints are enforced through a classical post-processing routine. This methodology enables the recovery of optimal delivery schedules, provided a sufficient number of samples is collected from the QPU to resolve the solution space. We benchmark the hybrid workflow through numerical emulations and demonstrate its effectiveness on Pasqal’s Fresnel QPU, reporting hardware experiments with configurations of up to 100 atoms.
\end{abstract}

\section{Introduction}

Quantum computing offers a promising alternative for tackling combinatorial optimization problems that are challenging to solve using purely classical computational resources~\cite{abbas2024challenges}. Many problems in this class can be encoded in a form suitable for quantum computers~\cite{lucas2014ising, glover2022qubo}, enabling the use of quantum or hybrid quantum-classical solvers~\cite{ajagekar2020hybrid}. Despite their recent breakthroughs in hardware technology, contemporary quantum processors have not yet reached full computational maturity and remain in the pre-fault-tolerant era~\cite{Preskill_2018,eisert2025mindgapsfraughtroad,DiVincenzo_2025}. Indeed, current quantum computers still suffer from noise, limited qubit connectivity, and short coherence times, making it crucial to assess which computational tasks can effectively benefit from near-term quantum hardware architectures~\cite{babbush2025grandchallengequantumapplications}.

Among other approaches to quantum computation, neutral atom–based devices~\cite{saffman2010rydberg, henriet2020neutralatoms, browaeys2020rydberg, bluvstein2021rydberg, Wintersperger2023}  are particularly effective for solving problems involving independent sets (ISs) of a graph~\cite{Dalyac2024,wurtz2024industry}. Indeed, the Rydberg blockade mechanism, induced by the dipole-dipole interaction~\cite{lukin2001dipole, urban2009rydberg}, can be leveraged in analog mode to energetically penalize configurations in which nearby atoms are simultaneously excited to the Rydberg state~\cite{Bernien2017}. By spatially arranging the atoms according to the desired graph topology, one can then generate excited-state configurations that represent ISs~\cite{serret2020solving}. This approach can be exploited to search the maximum independent set of large graphs~\cite{pichler2018optimization, ebadi2022quantumMIS,cazals2025identifyinghardnativeinstances,rava2025benchmarkingneutralatombasedquantum,7dkh-crjj}, a provably NP-hard computational problem~\cite{karp1972reducibility}, even for deceptively simple graph geometries~\cite{clark1990unit}. Importantly, neutral atom quantum computers can be leveraged to address a broader class of combinatorial optimization problems~\cite{wurtz2024industry,grotti2025practicalusecasesneutral}, including vertex coloring. Achieving this can require the integration of the quantum processing unit (QPU) into a hybrid quantum–classical workflow~\cite{graham2022rydberggraph, dasilva2023column}. In some of these hybrid schemes, the neutral atom QPU is used as a subroutine to generate ISs, which can then be inserted within classical algorithms such as branch-and-bound or column generation methods~\cite{vercellino2023bbqmis, vercellino2025hybrid,perron2025cg}.

In this work, we explore the use of neutral atom quantum computers to tackle an industrial problem in the field of logistics, the Drone Delivery Packing Problem (DDPP)~\cite{meng2023ddpp}. Specifically, we consider a delivery setting in which multiple drones, carried by a moving truck, distribute parcels to users scattered across a given territory. The objective of the DDPP is to determine the schedule that allows all deliveries to be completed using the minimum number of drones, hence reducing the operational costs and increasing the sustainability of the drone-based delivery systems. Mathematically, the optimization problem is characterized by constraints regarding delivery time windows and battery capacity of the drones~\cite{tarquini2024testing}. In the worst case, instances of this problem belong to the NP-hard complexity class~\cite{bettisorbelli2022greedy, 10.1145/3571306.3571411}. Overall, our main contributions are the following:
\begin{itemize}
    \item We reformulate the Drone Delivery Packing Problem as a partitioning problem based on independent sets of a scheduling graph encoding delivery incompatibilities; each IS corresponds to a time-feasible single-drone schedule; 
    \item We propose a hybrid quantum-classical pipeline that allows for the decoupling of the DDPP constraints: while the neutral-atom QPU is used as a stochastic sampler of single-drones schedules, a greedy classical correction routine enforces battery-budget feasibility, rendering the pool of candidate schedules feasible and suitable for the recombination in a graph partitioning workflow. Our method is based on the idea of leveraging the blockade mechanism to seek samples (time-feasible schedules) of a given size (number of deliveries) that are further classically corrected in terms of drones' battery.
    \item We benchmark the method using both Pasqal's Emulation \cite{Silv_rio_2022} and QPU \cite{henriet2020neutralatoms}: our analysis includes a study of the scaling with the dimension (number of deliveries) and with the number of samples, showing near-optimal performance on realistic instances of medium size. Particularly, our tests' size reaches up to the maximum 100-atom register for Pasqal's hardware.
\end{itemize}

\section{The Drone Delivery Packing Problem\!}\label{sec:ddpp}

The DDPP is a relevant problem in logistics that involves the optimal assignment of deliveries to a limited number of drones transported by a truck.  
The truck moves on a pre-defined route and serves as a mobile base station for the take-off and return of drones delivering parcels. The goal is to minimize the number of drones needed, under constraints of battery- and time-consistency. 
We consider a collection of identical drones $\mathcal{D}$, with a battery duration $B>0$. Let $\mathcal{N} = \{1, ..., N\}$ denote the set of deliveries to be completed and $c_j$ the energy cost required to complete the delivery $j \in \mathcal{N}$. The truck departs from the depot at a given time and follows its pre-defined route. To perform a delivery $j \in \mathcal{N}$, a drone launches from the truck at one of the designated locations along the truck route. After the completion of the delivery, the designated drone rejoins the vehicle. Let $t_j^L$ and $t_j^R$ represent the times at which the drone leaves and returns to the truck, respectively. Thus, the time interval $I_j = [t_j^L,t_j^R]$ specifies the flight duration to complete delivery $j$. Once all customers have received their parcels, the truck with the drones returns to the depot. 

The problem has been previously formulated as an Integer Linear Program (ILP) where binary variables $x_{ij}$ indicate whether drone $i$ is assigned to delivery $j$~\cite{10.1145/3571306.3571411}. The standard ILP formulation has a number of variables that scale with the number of deliveries and the upper bound on the number of drones $\mathcal{D}$; if the latter is conservatively set to $N$, the number of variables scales as $O(N^2)$. Although exact state-of-the-art classical solvers, such as branch-and-bound approaches~\cite{gurobi}, can obtain optimal solutions, their runtime deteriorates significantly for large instances, reflecting the asymptotic NP-hardness of the problem. The DDPP can also be recast into a Quadratic Unconstrained Binary Optimization (QUBO) problem~\cite{tarquini2024testing}, amenable to solution using classical and quantum annealing approaches~\cite{hauke2020annealing}. However, the QUBO formulation of the DDPP requires an even larger number of binary variables, due to the introduction of slack variables to encode inequality constraints.  

In this work, we adopt an alternative formulation that allows us to tackle the problem using neutral atom QPUs. To this aim, we introduce the \textit{scheduling graph} $G = (\mathcal{N}, E)$, where nodes correspond to deliveries and an edge $(i,j) \in E$ exists if and only if deliveries $i$ and $j$ overlap in time, i.e. are incompatible and cannot be assigned to the same drone. Within this setting, the DDPP can be seen as a minimum vertex coloring problem on $G$, in which the battery constraints limit the number of nodes that can be assigned to each color. To address this problem using neutral atoms, we build on prior works that recast the minimum vertex coloring problem as a \textit{partitioning problem} using the ISs of the graph~\cite{doi:10.1287/ijoc.8.4.344,dasilva2023column, vercellino2025hybrid}. Indeed, covering all nodes (deliveries) with the smallest possible number of disjoint ISs corresponds to minimum vertex coloring~\cite{doi:10.1287/ijoc.8.4.344,HANSEN2009135}. The DDPP can be formulated in the same way,  with each IS representing a set of deliveries assigned to a given drone. We note, however, that the set of valid ISs needs to be restricted to those that fulfill the battery constraint. 

Concretely, denoting by $\mathcal{I}$ the set of all ISs of the graph $G$, for each $s\in \mathcal{I}$ we introduce
\begin{align*}
    b_{js} &= \begin{cases}
        1, \ \text{$j \in \mathcal{N}$ belongs to the IS $s$ }\\
        0, \ \text{otherwise}
    \end{cases} \notag \\
    y_s &= \begin{cases}
        1, \ \text{$s$ is selected}\\
        0, \ \text{otherwise}
    \end{cases}
\end{align*}
In addition to $\mathcal{I}$, we also define the set of ISs that fulfill the battery constraint, namely 
\begin{equation}
    \mathcal{I}_B = \left\{ s \in \mathcal{I} \, \left| \, \sum\nolimits_{j \in \mathcal{N}} b_{js} c_j \leq B \right.\right\}. 
\end{equation}
The IS-based formulation of DDPP (IS-DDPP) corresponds to the following linear program
\begin{align} 
\min \ \ & \mathcal{D}= \sum_{s \in \mathcal{I}_B} y_s \label{eq:min-coloring-definition} \\ 
\text{s.t.} \ \ & \sum_{s \in \mathcal{I}_B} b_{js} \, y_s = 1, && \forall j \in \mathcal{N}, \label{eq:one-hot-const}\\ 
& y_s \in \{0,1\}, && \forall s \in \mathcal{I}_B. \label{eq:isddpp-vars} 
\end{align} 
Each IS corresponds to a set of deliveries that can be assigned to a single drone without time conflicts, and minimizing the number of disjoint ISs covering all vertices ($\mathcal{D}$) is equivalent to minimizing drone usage. Constraint~\eqref{eq:one-hot-const} ensures that each vertex is covered by a single selected IS, while the constraint on the battery is imposed by restricting the variables $s$ to $\mathcal{I}_B$.

The IS-DDPP in Eq.~\eqref{eq:min-coloring-definition} is not suitable for direct solution via branch-and-bound methods as the number of ISs of $G$ typically grows exponentially with the number of vertices $N$~\cite{https://doi.org/10.1002/jgt.3190110403, GRIGGS1988211}; thus, the task of enumerating all ISs is itself NP-hard in general~\cite{karp1972reducibility}. Even though $\mathcal{I}_B$ can contain fewer elements than $\mathcal{I}$, the number of variables $y_s$ can remain prohibitively large in practice, except for very small values of the battery $B$. For this reason, we adopt an approach in which we restrict IS-DDPP to a subset of ISs that respect the battery constraint, ${\mathcal{S}_B \subseteq \mathcal{I}_B}$, and then solve it by branch-and-bound~\cite{gurobi}. The set $\mathcal{S}_B$ is generated by using samples from a neutral atom QPU, corrected by a post-processing routine. 

Noticeably, we found that solving the IS-DDPP using the whole set $\mathcal{S}_B$ containing hundreds or a few thousand samples using state-of-the-art branch-and-bound solvers requires a few seconds on a laptop, and the bottleneck remains the IS sampling. As a consequence, we decided not to use more advanced schemes such as column generation to reduce the sample size to iteratively generate additional variables~\cite{dasilva2023column}, as in the tests we conducted on IS-DDPP, they led to solution quality degradation.

\section{Computational method}\label{sec:CG-not-decomposed}

\begin{figure}[t]
  \centering
  \includegraphics[width=\columnwidth]{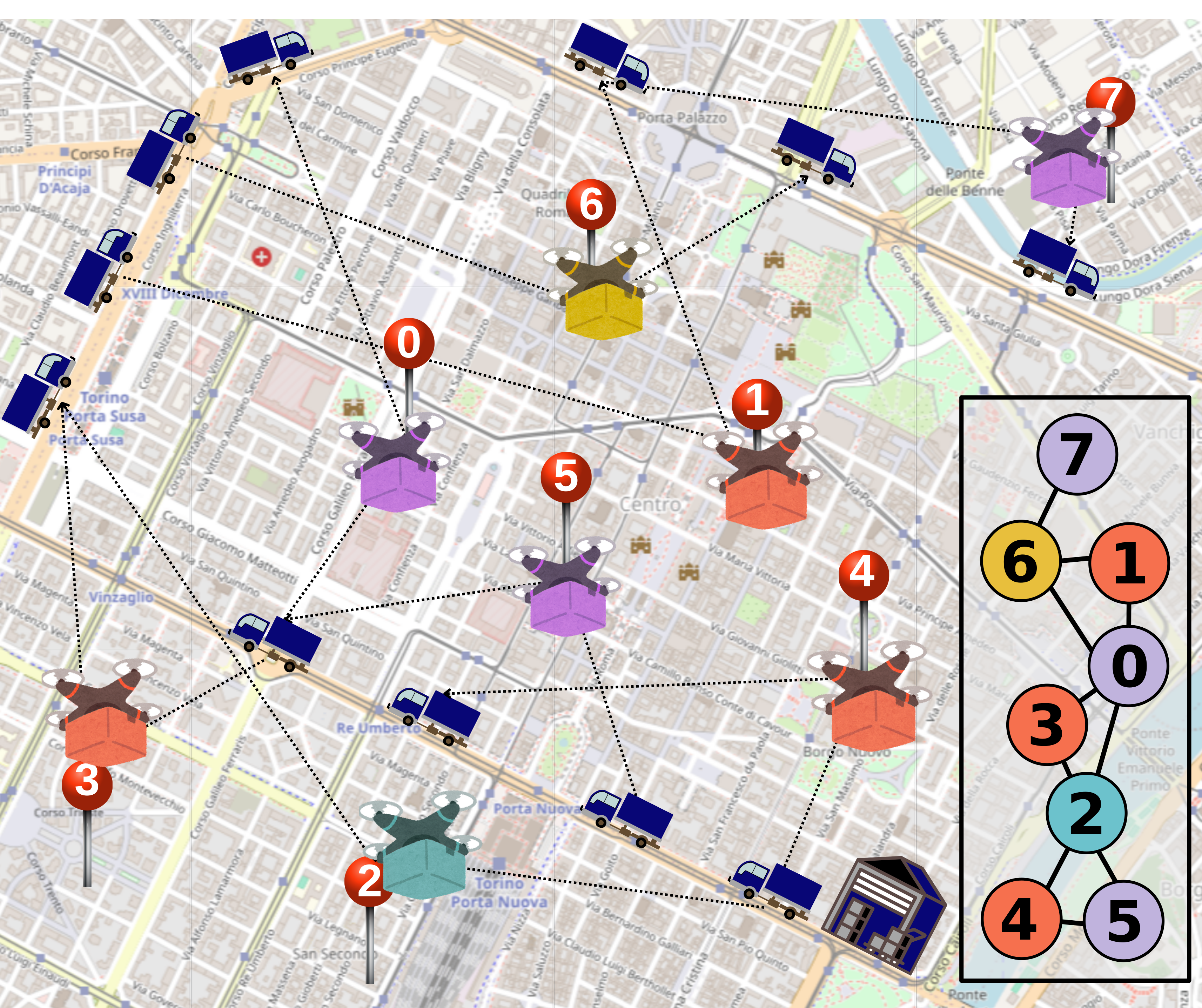}
    \caption{Delivery schedule for a DDPP instance with $N=8$ deliveries in the center of Turin, Italy. The DDPP is a (battery-constrained) minimum vertex coloring of the scheduling graph $\mathcal{G}$ shown in the inset, with edges indicating time-conflicting deliveries. The optimal solution is represented by nodes colors, which denote deliveries assigned to the same drone.
    }
\label{fig:8-dels-scheduled}
\end{figure}

\subsection{Neutral atom QPU for IS sampling}\label{sec:qpu_i_love_you}

In this section, we propose a simple quantum-classical scheme to tackle the IS-DDPP. For brevity, this approach is hereafter referred to as "QIS". As a first step, we use the neutral atom QPU as a sampler of ISs. Neutral atom analog computers can naturally encode the node independence constraint since, at low energy, the physical system is described by the Rydberg Hamiltonian
\begin{align}
    H(t) = \Omega(t)\sum_{i\in \mathcal{N}} \hat{\sigma}_i^x - \delta(t) \sum_{i\in \mathcal{N}} \hat{n}_i + \sum_{i<j \in \mathcal{N}} V_{ij} \hat{n}_i \hat{n}_j.
    \label{eq:rydberg-hamil}
\end{align}
Here, $\Omega(t)$ and $\delta(t)$ are the Rabi frequency and detuning parameters, respectively, which can be adjusted by tuning external pulses; $\hat{\sigma}_i^x$ is the first Pauli matrix acting on site $i$, and $\hat{n}_i$ is the number operator. The $V_{ij}$ term is the dipole-dipole interaction, which scales as $C_6/r_{ij}^6$ in the van der Waals regime, where $r_{ij}$ represents the distance between atoms $i$ and $j$. This interaction induces the Rydberg blockade, a mechanism that prevents atoms within a characteristic radius $R_{\rm Ryd} \approx \sqrt[6]{C_6/\Omega_{\rm max}}$ from being simultaneously excited~\cite{lukin2001dipole}, where $\Omega_{\rm max}$ represents the maximum Rabi frequency of the driving field. Within this radius, the energy shift due to the dipole-dipole interaction is strong enough to ensure that if one atom is in the $|1\rangle$ state, its neighbors are blocked from transitioning out of the ground state.

The Rydberg blockade provides a direct physical mechanism to encode the IS constraint on a graph. By mapping each node $j \in \mathcal{N}$ to an atom, the graph topology is realized through the atomic spatial arrangement: nodes connected by an edge are positioned within the blockade radius $R_{\rm Ryd}$ of one another~\cite{pichler2018optimization}. Then, leveraging the blockade constraint, one can exploit a quantum adiabatic algorithm~\cite{Albash_2018} to obtain ISs of properly encoded graphs with high probability~\cite{henriet2020neutralatoms}. At the end of a suitably designed adiabatic evolution, the measurement of the whole atomic register in the computational basis yields a bitstring $|\mathbf{n}\rangle = \bigotimes_{j \in \mathcal{N}} |n_j\rangle$ that represents a candidate IS, where node $j$ belongs ($n_j=1$) or does not belong ($n_j=0$) to the set. The size of the candidate IS is the Hamming weight of the bitstring, i.e., the expectation value of $\hat{w}=\sum_{j \in \mathcal{N}} \hat{n}_j$.

To implement the adiabatic algorithm, we employ pulse protocols to control the values of $\Omega(t)$ and $\delta(t)$, using the piecewise waveforms discussed in~\cite{Silv_rio_2022}, defined over three intervals. For the Rabi frequency, $\Omega(t)$ initially increases from $0$ to $\Omega_{\rm max}$ for a time $T_{\rm rise} = T/9$, then it remains constant at the value $\Omega_{\rm max}$ with duration $T_{\rm sweep} = 2T/3$, and finally it decreases to $0$ in time $T_{\rm fall} = 2T/9$ at the end of the sequence. Simultaneously, the detuning $\delta(t)$ is varied according to the same three stages with the following protocol: an initial plateau at constant value $\delta_{\rm min} = -3\Omega_{\rm max}$, a linear ramp and a final plateau at the final value $\delta_{\rm max}$. As discussed in \Cref{subsec:embed}, the value of $\Omega_{\rm max}$ is tied to the positions of the atoms and the Rydberg blockade radius, and is therefore fixed. On the other hand, $\delta_{\rm max}$ and $T$ need to be optimized. To this end, we tune their values to obtain a distribution of bitstrings with average Hamming weight $\langle \hat{w} \rangle$ close to (or slightly below) $\overline{w} = B N / (\sum_{j \in \mathcal{N}} c_j)$. The latter value represents the average delivery load per drone in a relaxed version of the problem, in which deliveries are treated as continuous quantities that can be partitioned to exactly fill the battery capacity of each drone. Our analysis indicates that this rule of thumb for the optimization of the parameters also yields the best results in terms of the final number of drones. 

We note, however, that at the end of the evolution protocol, due to hardware noise or non-adiabaticity, sampled bitstrings may still contain violations of the IS condition. Furthermore, sampled candidate ISs can also exceed the battery budget. Therefore, the samples obtained with neutral atoms require a classical post-processing to produce the set $\mathcal{S}_B$ of battery-feasible IS for IS-DDPP.

\begin{figure*}[t]
    \centering
    \includegraphics[width=0.9\linewidth]{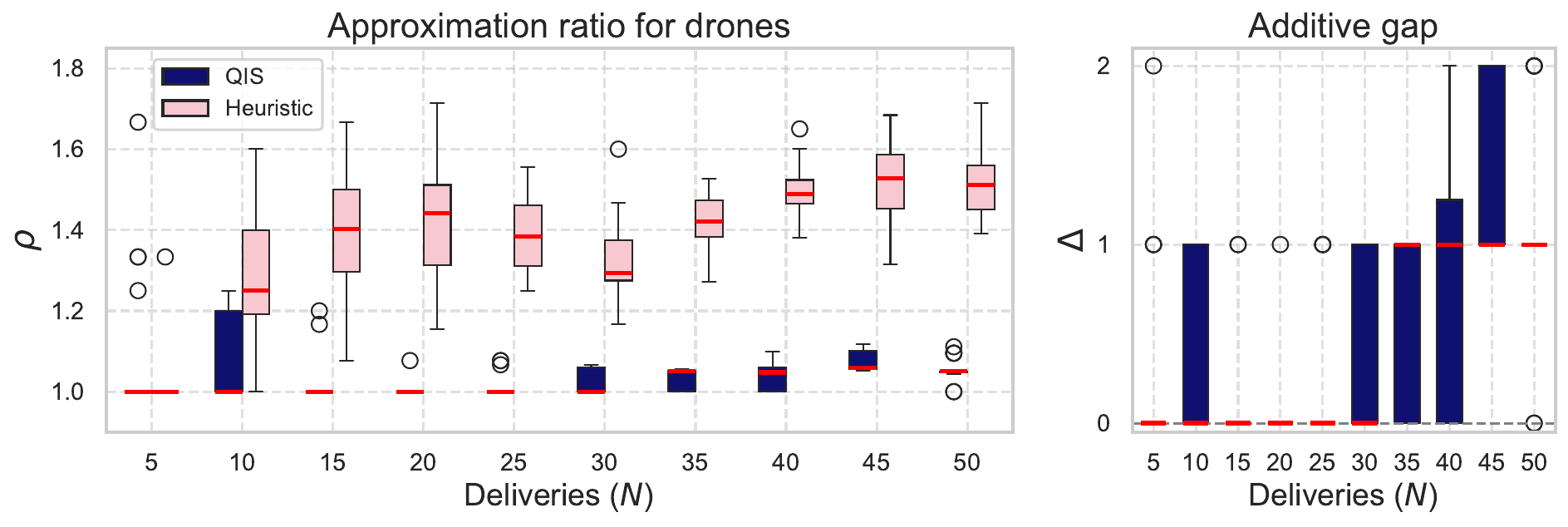}
    \caption{Box plots showing the approximation ratio $\rho$ (\textbf{left panel}) and additive gap $\Delta$ (\textbf{right panel}) for DDPP instances with $5$ to $50$ deliveries ($20$ instances per size), using Pasqal's emulator. $N_{\rm meas}$ is fixed to 500 samples.}
    \label{fig:approx-ratio-5-30}
\end{figure*}

\subsection{Post-processing routine}

We denote the collection of bitstrings obtained by the QPU as $\mathcal{S}_{\rm meas}$, with $N_{\rm meas}=|\mathcal{S}_{\rm meas}|$ being the total number of measurements. The following greedy post-processing routine is applied to the elements of $\mathcal{S}_{\rm meas}$:
\begin{enumerate}
\item \textbf{IS enforcement:} the algorithm identifies all nodes violating the IS condition and computes, for each of them, their conflict degrees (the number of neighbors each node has within the current set). Among conflicting nodes, it selects those with the highest conflict degree and removes one of them. When choosing the node to remove, the battery budget is considered. If the current set also exceeds the battery budget, the algorithm prioritizes the maximal-degree node with the highest cost, so that the removal contributes simultaneously to restoring feasibility with respect to both constraints. Otherwise, the node is chosen uniformly at random among the maximal-degree ones. This process iterates until no conflicting pairs remain, yielding an IS of $G$.
\item \textbf{Budget enforcement:} if the resulting IS still exceeds the battery budget $B$, vertices are iteratively removed to reduce total cost, prioritizing nodes with higher costs.
\end{enumerate}
This procedure yields a corrected collection $\mathcal{S}_B$ of time- and battery-feasible schedules for a single drone. To ensure that a feasible schedule is always returned, we also add the set of single-vertex ISs of the graph (\textit{singletons}), i.e. ${\{ \{j\}, \; j \in \mathcal{N}\}}$, to $\mathcal{S}_B$. This can then be used in IS-DDPP.

We note that the proposed approach is heuristic. Indeed, the neutral atom sampler is stochastic and may not generate the optimal ISs to ensure the exact partitioning of $G$, within a finite $N_{\rm meas}$. Furthermore, hardware effects can lead to violations of the Rydberg blockade. While the corrective post-processing enforces the feasibility of the solutions found, it does not guarantee optimality unless $\mathcal{S}_B = \mathcal{I}_B$. 

\Cref{fig:8-dels-scheduled} provides an example of an 8-delivery DDPP instance in the historic center of Turin, Italy (map from OpenStreetMap~\cite{openstreetmap}). The inset shows the associated scheduling graph, in which edges indicate time conflicts between deliveries. Nodes sharing the same color represent deliveries assigned to the same drone in the optimal schedule.

\subsection{Graph embedding on neutral atoms}\label{subsec:embed}

A primary challenge in utilizing neutral atom QPUs for graph problems is the physical \textit{embedding} of the graph topology onto a two-dimensional atomic array. Indeed, to tackle IS problems in analog mode, the target graph must be representable as a Unit Disk Graph (UDG), namely a geometric graph where edges between nodes exist if and only if the distance between the corresponding atoms is below $R_{\rm Ryd}$. Finding UDG representations for generic graphs has been proven to be theoretically NP-hard~\cite{BREU19983} and practically difficult~\cite{vercellino2022neural, de_Correc_2025}. Moreover, not all graphs admit a UDG representation. 

To perform the embedding of the scheduling graph, we adopt the Fruchterman-Reingold algorithm~\cite{https://doi.org/10.1002/spe.4380211102, graph-draw}, frequently used for this task~\cite{graham2022rydberggraph, dasilva2023column, manetsch2024tweezerarray6100highly}, as implemented in the NetworkX Python library~\cite{SciPyProceedings11}. This method is used to obtain a positional embedding for the nodes. We then verify that there exists a finite window between the maximal distance of connected pairs of nodes ($d_l$) and the minimal distance of disconnected ones ($d_r$). If that is the case, we perform a global rescaling of the nodes positions such that the radius $R_{\rm Ryd}$ coincides with the geometric mean of the aforementioned distances, i.e. $\sqrt{d_l d_r}$, consistent with prior works~\cite{PhysRevApplied.23.064023}. This ensures the graph topology is correctly embedded in the unit disk representation. 

We note that the embedding process presents several nontrivial challenges for practical use cases. The Fruchterman-Reingold algorithm is not guaranteed to yield a UDG representation for generic graphs. Furthermore, the final layout is subject to hardware constraints on the total area available for atom positioning and on the minimum interatomic distance. To address the latter constraint, one can change the effective blockade radius $R_{\rm Ryd}$ by tuning the value of $\Omega_{\rm max}$. Still, the problem of embedding generic graphs remains one of the primary bottlenecks for the use of neutral atom QPUs in industrial applications.

\section{Results}

\subsection{Benchmark using emulation on classical hardware}

We conduct numerical experiments to evaluate the performance of the proposed quantum-classical approach. For our numerical tests of the method in the emulation setting, we create a set of instances of the problem with a variable number of deliveries. 
 The following features define the DDPP instances generated for our experiments:
 \begin{itemize}
     \item Time windows (intervals): Each delivery $j$ is assigned a time window $I_j$, whose time duration is uniformly sampled within $5$ and $15$ minutes. These intervals are randomly distributed over a timespan of $7N$ minutes. We also ensure that each interval overlaps with at most $5$ others and that the resulting scheduling graph is not disconnected.
     \item Battery budget: Drones are assigned a battery life of $B \in \{30,40,50,60\}$ minutes. This reflects plausible flight endurance of delivery drones, which prioritize speed and robustness over long-duration operation~\cite{10.1108/IJLM-04-2023-0149, app12010207}.
     \item Costs: Each parcel $j$ has an associated cost
     \[
     c_j = W_j \times |I_j|
     \]
     where $W_j$ is a factor sampled from a log-uniform distribution in $(0,0.3)$, capturing the effect of varying parcel weights on energy consumption.
 \end{itemize}

\noindent To emulate the quantum adiabatic algorithm, we use the Pulser Python library developed by Pasqal~\cite{Silv_rio_2022}. For problem sizes below $20$ deliveries, we use the state-vector backend based on the QuTiP library~\cite{johansson2012qutip, johansson2013qutip}. For larger systems, we exploit the tensor-network-based backend~\cite{bidzhiev2025efficientemulationneutralatom}. The embedding of the graph is performed taking into account hardware constraints of the Fresnel QPU by Pasqal~\cite{henriet2020neutralatoms}. This architecture allows us to load atomic registers of size up to $100$ qubits.

Our numerical tests start with an assessment of the solution quality provided by the proposed workflow for IS-DDPP on small test cases. To this aim, we compare the proposed approach for DDPP with two classical methods. The first is the exact solution of the standard ILP formulation of DDPP~\cite{tarquini2024testing}, found with Gurobi, a state-of-the-art commercial solver (version 11.0)~\cite{gurobi}; the second is a simple heuristic method that combines the exact solution of the battery-unaware minimum vertex coloring, which is not NP-hard, followed by a post-processing step to ensure battery feasibility. Indeed, the solution of the minimum vertex coloring (ignoring battery constraints) on an interval graph can be obtained in polynomial time using the sequential \textit{greedy coloring} algorithm~\cite{GOLUMBIC1980171}. Following this, any drone assignment that violates battery constraints is recursively partitioned until all constraints are satisfied. 

The metrics for the comparison of solution quality are the approximation ratio $\rho$ and the additive gap $\Delta$, defined as 
\begin{align}
   &\rho = \frac{ \mathcal{D}}{\mathcal{D}_{\rm Exact}},\\
   &\Delta = \mathcal{D} - \mathcal{D}_{\rm Exact}
\end{align}
where $\mathcal{D}$ and $\mathcal{D}_{\rm Exact}$ are the numbers of drones obtained with an approximate method (QIS or heuristic methods) and the exact solver, respectively. An approximation $\rho \approx 1$ and $\Delta \approx 0$ indicates near-optimality relative to the exact solution. 

\Cref{fig:approx-ratio-5-30} summarizes the results of the tests on small- to medium-sized problems, ranging from $5$ to $50$ deliveries, with $20$ instances per size. The spatial embedding of each graph is performed by the aforementioned Fruchterman-Reingold algorithm. Multiple random initializations of the latter are often required to achieve a UDG representation that satisfies the atomic position constraints of Pasqal's hardware. The difficulty of finding a valid embedding scales significantly with the number of vertices, requiring a greater number of trials as the instance size increases. 
For QIS calculations, we sample $N_{\rm meas}=500$ bitstrings from the emulated quantum algorithm, which are then post-processed to yield $\mathcal{S}_B$.

\begin{figure}[t]
    \centering
    \includegraphics[width=1.0\linewidth]{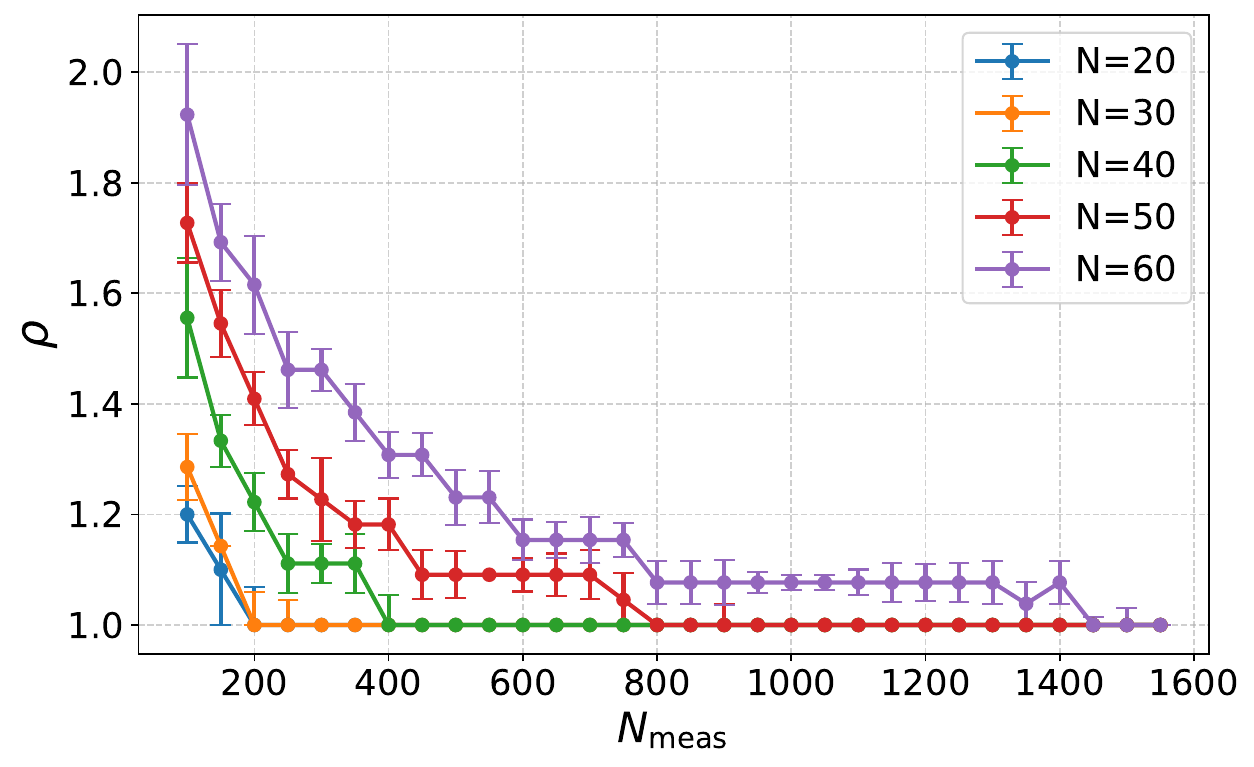}
    \caption{Approximation ratio as a function of the number of measurements $N_{\rm meas}$ for DDPP of different sizes $N$ (indicated by different colors). Data points correspond to the median value of $\rho$ obtained across 30 different sample sets, with relative error bars.}
    \label{fig:approx-vs-samples}
\end{figure}

The results of the left panel of \Cref{fig:approx-ratio-5-30} show that, across all sizes, the QIS method remains mostly concentrated near $\rho=1$, with only small deviations (typically within a few percent and up to $10\%$ above the exact solution at larger $N$). In contrast, the heuristic quality degrades with size: median ratios are substantially higher (about $\rho \approx 1.25$ at $N=10$, increasing to $\rho \approx 1.4$--$1.55$ for $N \ge 15$). They also exhibit larger dispersion and heavier upper tails. In the right panel of \Cref{fig:approx-ratio-5-30}, we show the value of $\Delta$, which is mostly $0$ and otherwise typically $1$, with worst cases of $2$ extra drones. Overall, the results indicate that the QIS pipeline systematically reduces the extra use of drones compared to the classical heuristic. 

The quality of the solutions obtained with QIS depends critically on the number of samples \(N_{\rm meas}\). This is evident in \Cref{fig:approx-ratio-5-30}, where keeping \(N_{\rm meas}=500\) fixed across all problem sizes leads to a progressive degradation of the gap \(\Delta\) as the instance size increases. To explicitly quantify the impact of sampling, in \Cref{fig:approx-vs-samples} we consider single instances at different sizes \(N=20,30,40,50,60\) and study the dependence of \(\rho\) on \(N_{\rm meas}\). For each instance, we vary \(N_{\rm meas}\) from 100 to 1600. We show that $\rho$ consistently improves by increasing $N_{\rm meas}$ until a threshold $N_{\rm meas}^*$ above which the QIS method is almost always able to retrieve the exact number of drones $\mathcal{D}_{\rm exact}$. A rough fit of this threshold vs the problem size $N$ leads numerically to an exponential scaling $N_{\rm meas}^* \sim O(1.07^N)$. This is compatible with the fact that the number of samples required to find the exact solution should well approximate $\mathcal{I}_B$, which generally contains an exponential number of ISs as $N$ increases.

\begin{figure}[ht]
    \centering
    \begin{subfigure}[t]{0.48\textwidth}
        \centering
        \caption{Sampled Bitstrings}
        \includegraphics[width=\textwidth]{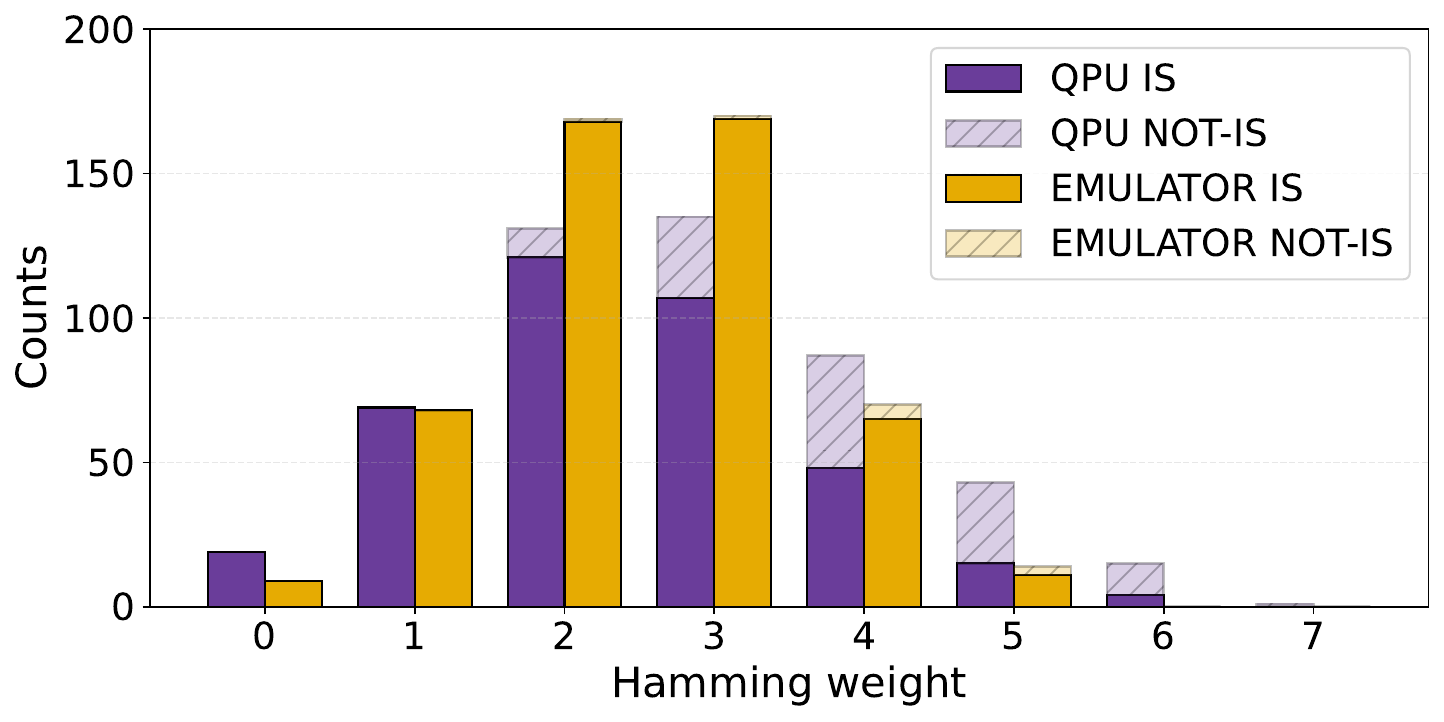}
    \end{subfigure}
    \hfill
    \begin{subfigure}[t]{0.48\textwidth}
        \centering
        \caption{Corrected Bitstrings}
        \includegraphics[width=\textwidth]{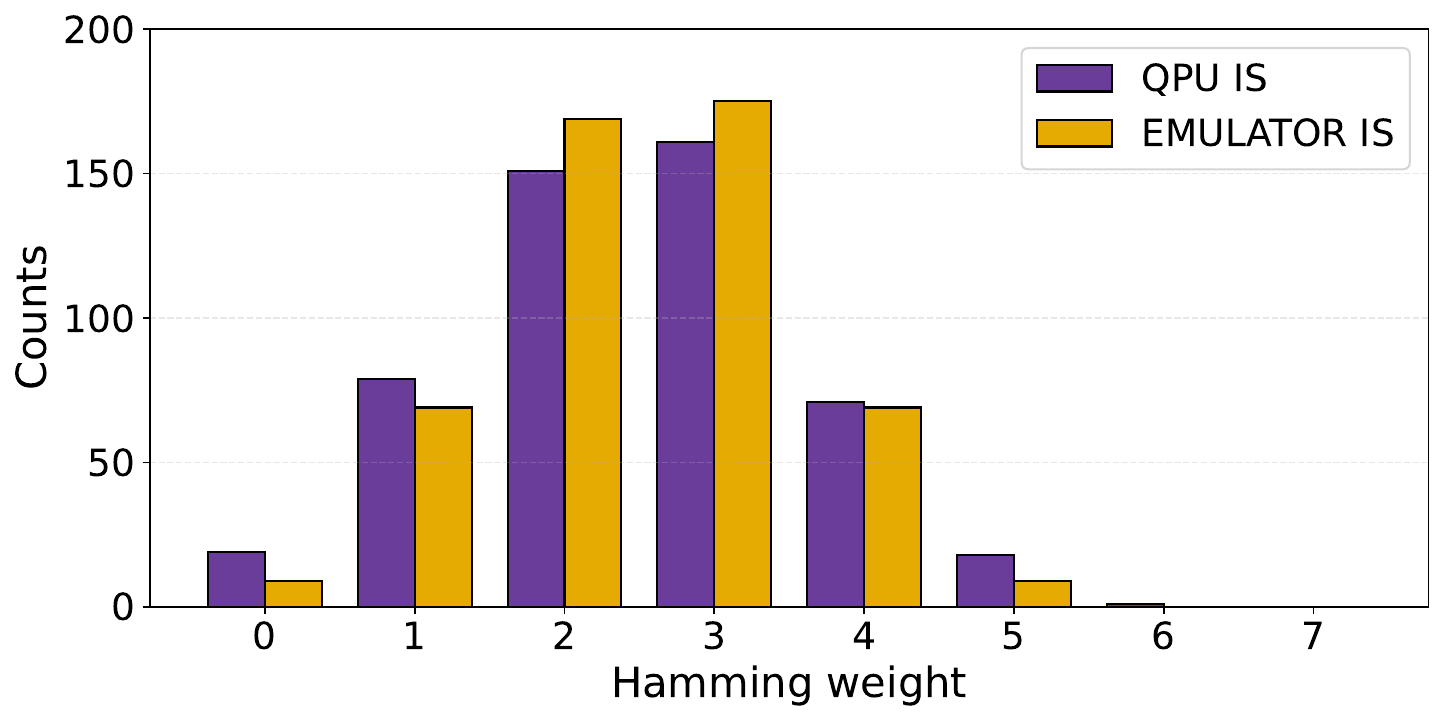}
    \end{subfigure}
        \caption{Results for the 20-delivery DDPP instance used in the experiments on Pasqal's QPU. \textbf{Top panel}: histogram showing the Hamming weight distribution of the neutral atom samples. Purple and orange bars represent results obtained on the QPU and by emulation, respectively. The solid (hatched) portions of the bars denote the number of samples that are (are not) valid IS of the scheduling graph. \textbf{Bottom panel:} analogous histogram for the post-processed samples forming the $\mathcal{S}_B$ set.}
    \label{fig:emulation-vs-QPU-20}
\end{figure}

\subsection{Small-instance test on a neutral atom QPU}

We proceed by describing the results of the experiments on Fresnel, the neutral atom QPU by Pasqal~\cite{henriet2020neutralatoms}. For these tests, we choose to fix the drone battery to $B=90$. We first consider a small instance with 20 deliveries, for which we collect $N_{\rm meas} = 500$ samples from the QPU, as well as from an emulation using Pulser.

The top panel of \Cref{fig:emulation-vs-QPU-20} shows the distribution of the Hamming weight of the neutral atom samples forming $\mathcal{S}_{\rm meas}$. A similar histogram is shown in the bottom panel for the corrected set $\mathcal{S}_B$ obtained by post-processing. Purple bars indicate the result on the real QPU, while orange bars indicate results obtained by emulation with Pulser. This figure emphasizes the similarity between the corrected QPU and emulation histograms for these system sizes, demonstrating that the post-processing effectively compensates for QPU noise (and violations of the battery constraint). For this relatively small test case, our hybrid algorithm correctly identifies the exact solution with $\mathcal{D}_{Exact} = 5$ drones both in emulation and on the Fresnel QPU. This result is important because it shows that the proposed quantum-classical pipeline is robust: although raw QPU samples and emulated samples may differ, the correction routine enables the use of battery- and time- feasible candidate schedules for the subsequent covering workflow.
In the emulation setting, the chosen $\delta_{\rm max}$ allows us to sample mostly battery-feasible ISs. Therefore, only a limited number of bitstrings require post-processing corrections, indicating that violations of independence or budget constraints occur only sporadically in the sample set for this problem instance of size 20. This suggests that the optimization of the distribution parameters is effective in this case. Concerning the QPU results, the effect of the correction routine is slightly more pronounced. Indeed, the QPU displays a higher number of IS violations to be corrected in the post-processing phase, probably due to the fact that the evolution time $T$ in practice is far from adiabatic. This figure demonstrates the dual role of the correction routine: removing battery infeasibility and mitigating errors, especially induced by hardware noise.

\subsection{Large-scale DDPP instances}

\begin{figure}[ht]
    \includegraphics[width=0.5\textwidth]{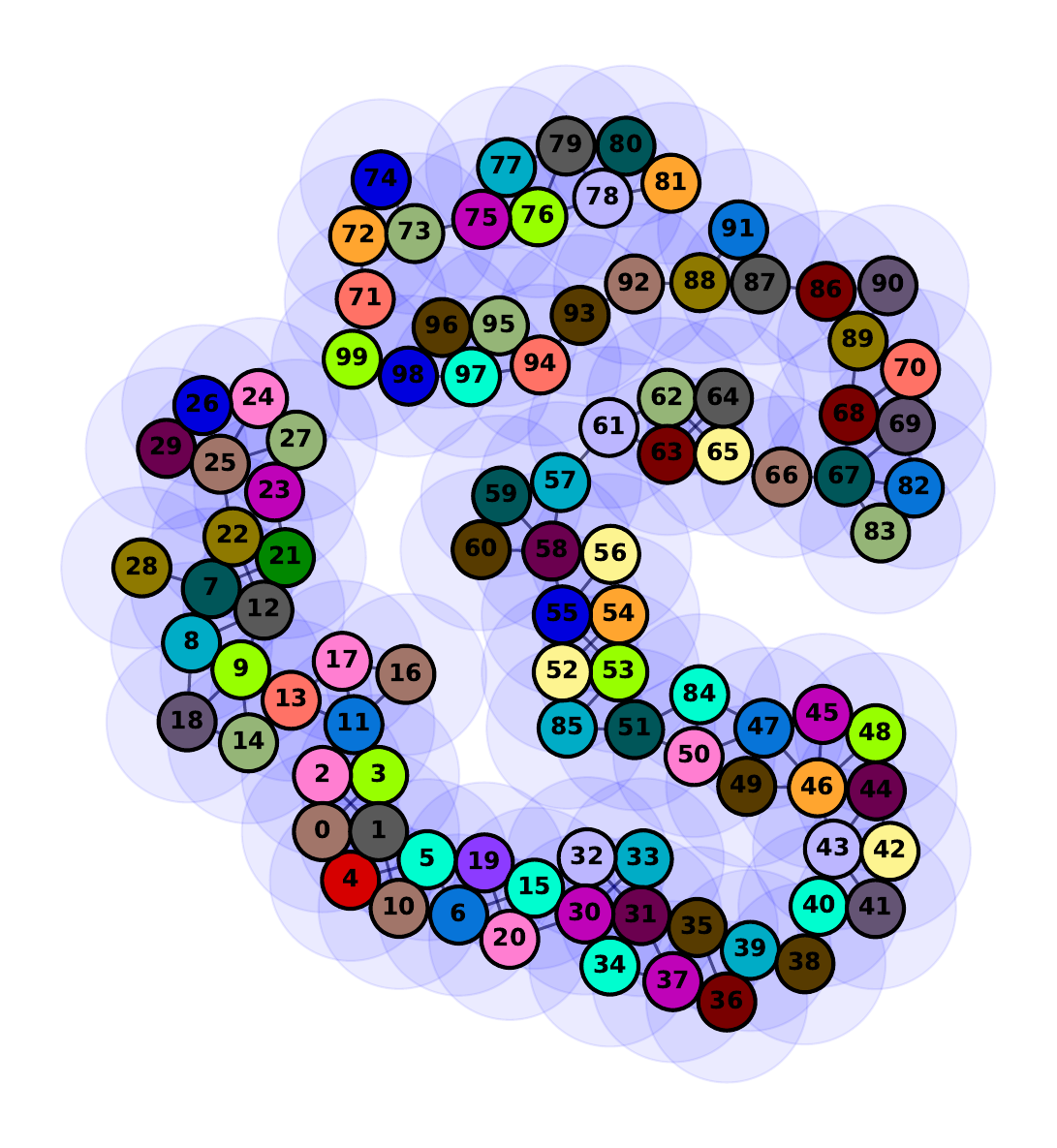}
    \caption{Scheduling graph of 100 deliveries used in the experiments on Pasqal's QPU. Nodes denote individual deliveries, while edges represent temporal conflicts. The coloring indicates the assignment of deliveries to specific drones, illustrating the solution to the battery-constrained graph coloring problem.}
    \label{fig:large-tested-graphs}
\end{figure}

\begin{figure}[ht]
    \includegraphics[width=0.5\textwidth]{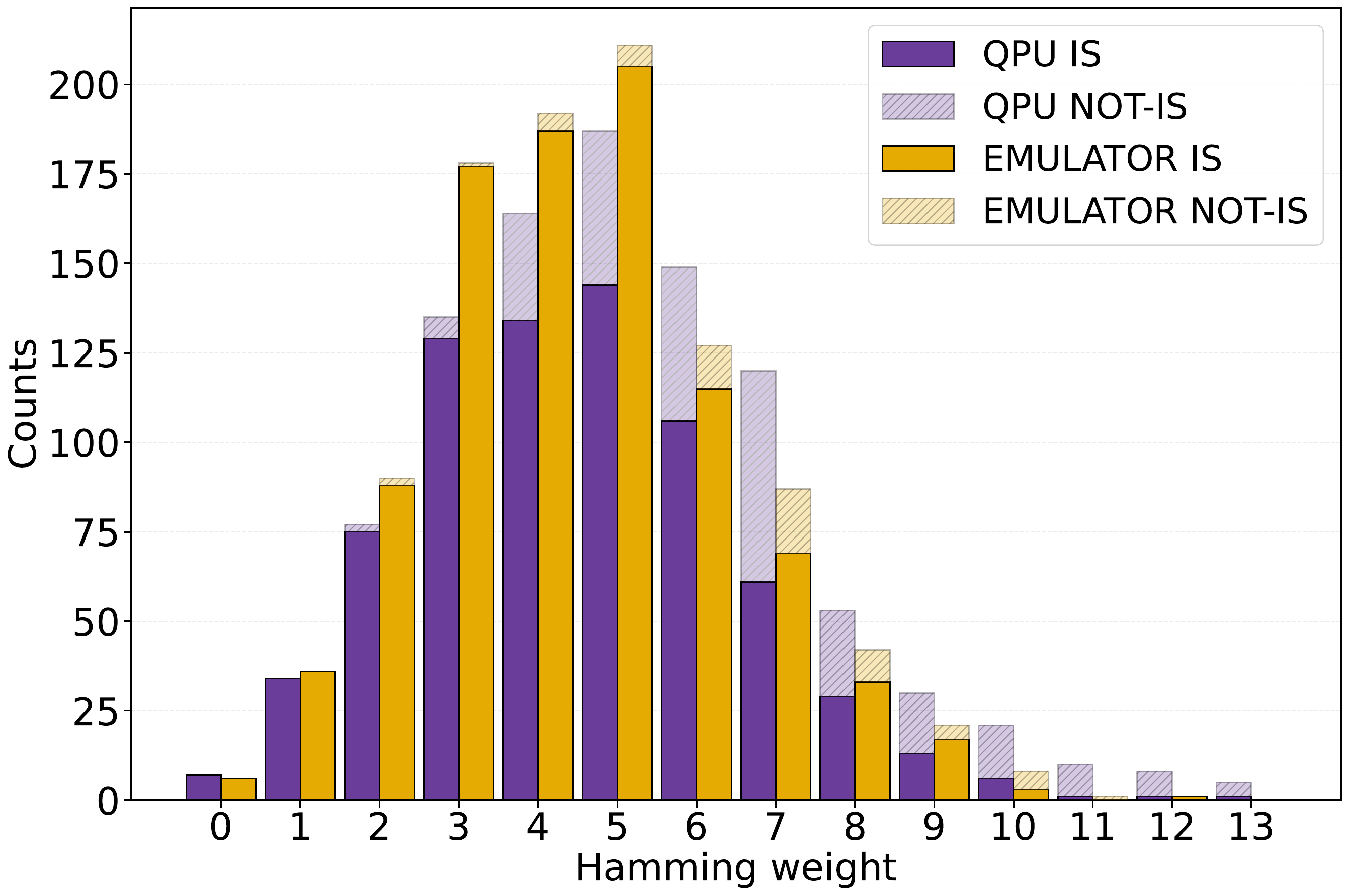}
        \caption{Hamming weight distribution of samples obtained via emulation (orange) and QPU (purple) for the 100-delivery instance shown in \Cref{fig:large-tested-graphs}. In both cases, $N_{\rm meas}=1000$ samples are collected. The solid (hatched) portions of the bars denote the number of samples that are (are not) valid IS of the scheduling graph. The percentage of valid IS among measured samples is 93.7\% in emulation and 74.1\% on the QPU.}
    \label{fig:QPU-vs-EMU-100}
\end{figure}

Here, we scale up the dimension of the problem instances tested on the QPU by generating instances with 50, 80 and 100 deliveries. While the instance with 20 atoms could be easily embedded automatically using the method described above, embedding instances with 50-100 atoms proved to be practically infeasible. Therefore, to still be able to run tests, we constructed a valid delivery graph by manually placing the atoms using the Pulser web interface. This highlights the difficulty of embedding medium/large graphs onto the neutral atom QPU, while simultaneously meeting the competing requirements of minimum atomic distance, constraint on the dimension of the register area, and the unit-disk property imposed by the Rydberg blockade mechanism. In these experiments, the maximum amplitude $\Omega_{\rm max}$ was fixed at $2 \pi \ \mathrm{rad}/\mu\mathrm{s}$. \Cref{fig:large-tested-graphs} displays the scheduling graph associated with the tested instance of size 100, with the best coloring obtained with our method. Nodes represent deliveries, and their positions coincide with the layout assigned to the atoms on the QPU. The shaded circles indicate $R_{\rm Ryd}$ and highlight the UDG property of the graph. 

The remaining parameters describing the evolution schedule, i.e. the final detuning $\delta_{\rm max}$ and the total time $T$, were optimized over a finite grid such that the average Hamming weight of the samples roughly matches the value $\overline{w}$, as discussed in \Cref{sec:qpu_i_love_you}. To obtain samples with the correct weight range, we choose a relatively short pulse schedule of duration $T=450$ ns, with $\delta_{\rm max} \approx -0.5\Omega_{\rm max}$. This value can be explained in light of the non-adiabaticity of the pulse schedule. Indeed, we notice that, in a strictly adiabatic regime, $0 < \delta_{\rm max} < \min_{(i,j) \in E} V_{ij}$ would favor the maximum IS, while $\delta_{\rm max} < 0$ would push the system towards a state with all the atoms in the $|0\rangle$ state.
In either case, the resulting Hamming weight distribution would deviate from the target value $\overline{w} \approx 5$ required for the IS-DDPP instances studied here. Overall, this indicates that, while adiabatic schedules are suited to find the maximum IS, they cannot be effectively used to sample ISs with an intermediate Hamming weight. Nevertheless, working away from the adiabatic regime with relatively short pulses allows us to carefully optimize $\delta_{\rm max}$ values and select ISs in the correct regime. In our tests up to 100 deliveries, we observe a smooth monotonic dependence of the average Hamming weight $\langle \hat{w} \rangle$ with respect to $\delta_{\rm max}$, so that the calibration of this parameter can be done straightforwardly. Additionally, the optimal parameters describing the schedule appear to be fairly generalizable to problems with similar sizes. Indeed, we find similar values of optimal $\delta_{\rm max}$ and $T$ for $N=50$, $N=80$ and $N=100$.

\Cref{fig:QPU-vs-EMU-100} presents the Hamming weight distribution of samples obtained from the 100-delivery instance of \Cref{fig:large-tested-graphs} using the optimized pulse schedule. The sample distribution obtained from the Fresnel QPU closely resembles the emulator results, with the main discrepancy being a higher number of IS violations, likely attributable to device noise and other non-ideal hardware effects. 
For this specific problem instance, the quantum method with $N_{\rm meas}=1000$ samples yields a final solution requiring $\mathcal{D}_{\rm QIS}=23$ drones, to be compared with the exact value $\mathcal{D}_{\rm Exact} = 19$.

In \Cref{tab:placeholder} we summarize the results obtained on the Fresnel QPU with our method as compared to the exact minimum number of drones for varying number of deliveries $N$. In all the cases shown there, the emulated and the QPU-based sampling led to the same number of drones.
Specifically, the method achieves $\rho=1$ for the smallest considered instance with $N=20$, indicating that the QPU samples all the required ISs to find the optimal partitioning in that case. The approximation ratio $\rho$ deviates more and more from the ideal result as the size $N$ is increased. This is consistent with the results predicted in emulation and reflects the need for larger $N_{\rm meas}$ as the instance size increases. Still, $\rho$ maintains values close to 1, validating the potential of the proposed approach. 

\begin{table}[t]
    \centering
    \begin{tabular}{|c|c|c|c|}
    \hline
      Deliveries ($N$)  & $\mathcal{D}_{\rm Exact}$ & $\mathcal{D}_{\rm QIS}$ & $\rho$ \\
        \hline
      20 & 5 & 5 & 1.0 \\
      50 & 11 & 12 & 1.09\\
       80 & 16 & 18 & 1.12\\
       100 & 19 & 23 & 1.21\\
       \hline
    \end{tabular}
    \caption{Comparison of the number of drones obtained by the proposed QIS method (using Fresnel QPU) and exact results. We collected $N_{\rm meas}=500$ samples for $N=20,50$ and $N_{\rm meas}=1000$ samples for $N=80,100$.}
    \label{tab:placeholder}
\end{table}

\section{Conclusions}

In this work, we leveraged the capabilities of a neutral atom QPU within a hybrid algorithm to address the NP-hard Drone Delivery Packing Problem. We ran a quantum adiabatic algorithm on the QPU to sample candidate conflict-free drone assignments. The samples obtained this way are then corrected to satisfy the drone battery constraint using a classical subroutine. This scheme allows us to decouple the two primary DDPP constraints (time consistency and battery feasibility). In contrast to previous quantum approaches based on Quadratic Unconstrained Binary Optimization (QUBO), which frequently yield infeasible schedules due to penalty violations \cite{tarquini2024testing}, our hybrid scheme consistently generates feasible solutions, even for large-scale delivery problems approaching industrially relevant size. Nevertheless, our results indicate that further developments are required to make this approach competitive with state-of-the-art commercial solvers.

We tested the hybrid method by numerical emulation on different instances with varying time windows, costs, and drone endurance. The results show that high solution qualities can be achieved up to 50 deliveries with a reasonable number of measurements. We also show that this hybrid approach is systematically improvable by increasing the number of QPU samples, but the exact solution is generally achieved only with an exponential number of measurements. 

Additionally, we performed experiments using Pasqal’s Fresnel neutral atom QPU, tackling DDPP instances with up to 100 deliveries, corresponding to the maximum number of available qubits. While the QPU exhibits a higher frequency of IS violations compared to the emulation, the overall solution quality remains consistent with the emulator predictions. This suggests that our framework is fairly robust to hardware noise, provided that the corrected subset $\mathcal{S}_B$ adequately covers the scheduling graph.

While our work demonstrates that problems of this kind can be addressed within hybrid quantum–classical frameworks that enforce feasibility, it also highlights the practical challenges of encoding real-world instances on a neutral atom QPU, arising from the unit-disk constraint and the finite size of the addressable region of the machine. In particular, previously proposed embedding methods were unable to automatically embed the larger instances with 50-100 atoms, requiring manual tuning of the atomic positions. This constitutes a major operational bottleneck in the deployment of these processors for general-purpose combinatorial optimization

Beyond the quantum strategy proposed here, our results motivate the development of efficient IS-based, quantum-inspired heuristics for the DDPP, representing a viable avenue for future research.

\section*{Authors contributions}
S.T. defined the mathematical formulation, implemented the methodology, developed the code, and performed all calculations (emulations and QPU runs). M.V. and F.F. contributed to the data analysis, manuscript writing, and figure preparation. D.D. and F.T. supervised the work. All authors contributed to editing the manuscript.

\section*{Acknowledgments}
We acknowledge a CINECA award under the ISCRA initiative for providing access to quantum computing resources. S.T.\ wishes to thank the Leonardo Quantum Computing Solutions group in Genoa for their assistance and collaboration throughout this project. F.T. is partially funded by the PRIN-MUR project MOLE (code 2022ZK5ME7) and by the PRIN-PNRR project FIN4GEO (code P2022BNB97).

\bibliographystyle{quantum}
\bibliography{lit} 

\begin{thebibliography}{10}

\bibitem{abbas2024challenges}
Amira Abbas, Andris Ambainis, Brandon Augustino, Andreas B{\"a}rtschi, Harry Buhrman, Carleton Coffrin, Giorgio Cortiana, Vedran Dunjko, Daniel~J. Egger, Bruce~G. Elmegreen, Nicola Franco, Filippo Fratini, Bryce Fuller, Julien Gacon, Constantin Gonciulea, Sander Gribling, Swati Gupta, Stuart Hadfield, Raoul Heese, Gerhard Kircher, Thomas Kleinert, Thorsten Koch, Georgios Korpas, Steve Lenk, Jakub Marecek, Vanio Markov, Guglielmo Mazzola, Stefano Mensa, Naeimeh Mohseni, Giacomo Nannicini, Corey O'Meara, Elena Pe{\~n}a~Tapia, Sebastian Pokutta, Manuel Proissl, Patrick Rebentrost, Emre Sahin, Benjamin C.~B. Symons, Sabine Tornow, V{\'\i}ctor Valls, Stefan Woerner, Mira~L. Wolf-Bauwens, Jon Yard, Sheir Yarkoni, Dirk Zechiel, Sergiy Zhuk, and Christa Zoufal.
\newblock ``Challenges and opportunities in quantum optimization''.
\newblock \href{https://dx.doi.org/10.1038/s42254-024-00770-9}{Nature Reviews Physics {\bf 6}, 718--735}~(2024).

\bibitem{lucas2014ising}
Andrew Lucas.
\newblock ``Ising formulations of many {NP} problems''.
\newblock \href{https://dx.doi.org/10.3389/fphy.2014.00005}{Frontiers in Physics{\bf 2}}~(2014).

\bibitem{glover2022qubo}
Fred Glover, Gary Kochenberger, Rick Hennig, and Yu~Du.
\newblock ``Quantum bridge analytics i: a tutorial on formulating and using qubo models''.
\newblock \href{https://dx.doi.org/10.1007/s10479-022-04634-2}{Annals of Operations Research {\bf 314}, 141--183}~(2022).

\bibitem{ajagekar2020hybrid}
Akshay Ajagekar, Travis Humble, and Fengqi You.
\newblock ``Quantum computing based hybrid solution strategies for large-scale discrete-continuous optimization problems''.
\newblock \href{https://dx.doi.org/10.1016/j.compchemeng.2019.106630}{Computers amp; Chemical Engineering {\bf 132}, 106630}~(2020).

\bibitem{Preskill_2018}
John Preskill.
\newblock ``Quantum computing in the nisq era and beyond''.
\newblock \href{https://dx.doi.org/10.22331/q-2018-08-06-79}{Quantum {\bf 2}, 79}~(2018).

\bibitem{eisert2025mindgapsfraughtroad}
Jens Eisert and John Preskill.
\newblock ``Mind the gaps: The fraught road to quantum advantage''~(2025).
\newblock  \href{http://arxiv.org/abs/2510.19928}{arXiv:2510.19928}.

\bibitem{DiVincenzo_2025}
David~P DiVincenzo.
\newblock ``Thirty years of quantum computing''.
\newblock \href{https://dx.doi.org/10.1088/2058-9565/ade0e4}{Quantum Science and Technology {\bf 10}, 030501}~(2025).

\bibitem{babbush2025grandchallengequantumapplications}
Ryan Babbush, Robbie King, Sergio Boixo, William Huggins, Tanuj Khattar, Guang~Hao Low, Jarrod~R. McClean, Thomas O'Brien, and Nicholas~C. Rubin.
\newblock ``The grand challenge of quantum applications''~(2025).
\newblock  \href{http://arxiv.org/abs/2511.09124}{arXiv:2511.09124}.

\bibitem{saffman2010rydberg}
M.~Saffman, T.~G. Walker, and K.~M\o{}lmer.
\newblock ``Quantum information with rydberg atoms''.
\newblock \href{https://dx.doi.org/10.1103/RevModPhys.82.2313}{Rev. Mod. Phys. {\bf 82}, 2313--2363}~(2010).

\bibitem{henriet2020neutralatoms}
Loïc Henriet, Lucas Beguin, Adrien Signoles, Thierry Lahaye, Antoine Browaeys, Georges-Olivier Reymond, and Christophe Jurczak.
\newblock ``Quantum computing with neutral atoms''.
\newblock \href{https://dx.doi.org/10.22331/q-2020-09-21-327}{Quantum {\bf 4}, 327}~(2020).

\bibitem{browaeys2020rydberg}
Antoine Browaeys and Thierry Lahaye.
\newblock ``Many-body physics with individually controlled {R}ydberg atoms''.
\newblock \href{https://dx.doi.org/10.1038/s41567-019-0733-z}{Nature Physics {\bf 16}, 132--142}~(2020).

\bibitem{bluvstein2021rydberg}
D.~Bluvstein, A.~Omran, H.~Levine, A.~Keesling, G.~Semeghini, S.~Ebadi, T.~T. Wang, A.~A. Michailidis, N.~Maskara, W.~W. Ho, S.~Choi, M.~Serbyn, M.~Greiner, V.~Vuletić, and M.~D. Lukin.
\newblock ``Controlling quantum many-body dynamics in driven rydberg atom arrays''.
\newblock \href{https://dx.doi.org/10.1126/science.abg2530}{Science {\bf 371}, 1355--1359}~(2021).

\bibitem{Wintersperger2023}
Karen Wintersperger, Florian Dommert, Thomas Ehmer, Andrey Hoursanov, Johannes Klepsch, Wolfgang Mauerer, Georg Reuber, Thomas Strohm, Ming Yin, and Sebastian Luber.
\newblock ``Neutral atom quantum computing hardware: performance and end-user perspective''.
\newblock \href{https://dx.doi.org/10.1140/epjqt/s40507-023-00190-1}{EPJ Quantum Technology {\bf 10}, 32}~(2023).

\bibitem{Dalyac2024}
Constantin Dalyac, Lucas Leclerc, Louis Vignoli, Mehdi Djellabi, Wesley da~Silva Coelho, Bruno Ximenez, Alexandre Dareau, Davide Dreon, Vincent~E. Elfving, Adrien Signoles, Louis-Paul Henry, and Lo{\"i}c Henriet.
\newblock ``Graph algorithms with neutral atom quantum processors''.
\newblock \href{https://dx.doi.org/10.1140/epja/s10050-024-01385-5}{The European Physical Journal A {\bf 60}, 177}~(2024).

\bibitem{wurtz2024industry}
Jonathan Wurtz, Pedro L.~S. Lopes, Christoph Gorgulla, Nathan Gemelke, Alexander Keesling, and Shengtao Wang.
\newblock ``{Industry applications of neutral-atom quantum computing solving independent set problems}''~(2022).
\newblock  \href{http://arxiv.org/abs/2205.08500}{arXiv:2205.08500}.

\bibitem{lukin2001dipole}
M.~D. Lukin, M.~Fleischhauer, R.~Cote, L.~M. Duan, D.~Jaksch, J.~I. Cirac, and P.~Zoller.
\newblock ``Dipole blockade and quantum information processing in mesoscopic atomic ensembles''.
\newblock \href{https://dx.doi.org/10.1103/PhysRevLett.87.037901}{Phys. Rev. Lett. {\bf 87}, 037901}~(2001).

\bibitem{urban2009rydberg}
E.~Urban, T.~A. Johnson, T.~Henage, L.~Isenhower, D.~D. Yavuz, T.~G. Walker, and M.~Saffman.
\newblock ``Observation of {R}ydberg blockade between two atoms''.
\newblock \href{https://dx.doi.org/10.1038/nphys1178}{Nature Physics {\bf 5}, 110--114}~(2009).

\bibitem{Bernien2017}
Hannes Bernien, Sylvain Schwartz, Alexander Keesling, Harry Levine, Ahmed Omran, Hannes Pichler, Soonwon Choi, Alexander~S. Zibrov, Manuel Endres, Markus Greiner, Vladan Vuleti{\'{c}}, and Mikhail~D. Lukin.
\newblock ``Probing many-body dynamics on a 51-atom quantum simulator''.
\newblock \href{https://dx.doi.org/10.1038/nature24622}{Nature {\bf 551}, 579--584}~(2017).

\bibitem{serret2020solving}
Michel~Fabrice Serret, Bertrand Marchand, and Thomas Ayral.
\newblock ``Solving optimization problems with rydberg analog quantum computers: Realistic requirements for quantum advantage using noisy simulation and classical benchmarks''.
\newblock \href{https://dx.doi.org/10.1103/PhysRevA.102.052617}{Phys. Rev. A {\bf 102}, 052617}~(2020).

\bibitem{pichler2018optimization}
Hannes Pichler, Sheng-Tao Wang, Leo Zhou, Soonwon Choi, and Mikhail~D. Lukin.
\newblock ``Quantum optimization for maximum independent set using rydberg atom arrays''~(2018).
\newblock  \href{http://arxiv.org/abs/1808.10816}{arXiv:1808.10816}.

\bibitem{ebadi2022quantumMIS}
S.~Ebadi, A.~Keesling, M.~Cain, T.~T. Wang, H.~Levine, D.~Bluvstein, G.~Semeghini, A.~Omran, J.-G. Liu, R.~Samajdar, X.-Z. Luo, B.~Nash, X.~Gao, B.~Barak, E.~Farhi, S.~Sachdev, N.~Gemelke, L.~Zhou, S.~Choi, H.~Pichler, S.-T. Wang, M.~Greiner, V.~Vuletić, and M.~D. Lukin.
\newblock ``Quantum optimization of maximum independent set using rydberg atom arrays''.
\newblock \href{https://dx.doi.org/10.1126/science.abo6587}{Science {\bf 376}, 1209--1215}~(2022).

\bibitem{cazals2025identifyinghardnativeinstances}
Pierre Cazals, Aymeric François, Loïc Henriet, Lucas Leclerc, Malory Marin, Yassine Naghmouchi, Wesley da~Silva~Coelho, Florian Sikora, Vittorio Vitale, Rémi Watrigant, Monique~Witt Garzillo, and Constantin Dalyac.
\newblock ``Identifying hard native instances for the maximum independent set problem on neutral atoms quantum processors''~(2025).
\newblock  \href{http://arxiv.org/abs/2502.04291}{arXiv:2502.04291}.

\bibitem{rava2025benchmarkingneutralatombasedquantum}
Andrea~B. Rava, Kristel Michielsen, and J.~A. Montanez-Barrera.
\newblock ``Benchmarking neutral atom-based quantum processors at scale''~(2025).
\newblock  \href{http://arxiv.org/abs/2511.22967}{arXiv:2511.22967}.

\bibitem{7dkh-crjj}
Martin J.~A. Schuetz, Ruben~S. Andrist, Grant Salton, Romina Yalovetzky, Rudy Raymond, Yue Sun, Atithi Acharya, Shouvanik Chakrabarti, Marco Pistoia, and Helmut~G. Katzgraber.
\newblock ``Quantum compilation toolkit for rydberg atom arrays with implications for problem hardness and quantum speedups''.
\newblock \href{https://dx.doi.org/10.1103/7dkh-crjj}{Phys. Rev. Res. {\bf 7}, 033107}~(2025).

\bibitem{karp1972reducibility}
Richard~M. Karp.
\newblock ``Reducibility among combinatorial problems''.
\newblock In Raymond~E. Miller, James~W. Thatcher, and Jean~D. Bohlinger, editors, Complexity of Computer Computations: Proceedings of a symposium on the Complexity of Computer Computations, held March 20--22, 1972, at the IBM Thomas J. Watson Research Center, Yorktown Heights, New York, and sponsored by the Office of Naval Research, Mathematics Program, IBM World Trade Corporation, and the IBM Research Mathematical Sciences Department.
\newblock \href{https://dx.doi.org/10.1007/978-1-4684-2001-2_9}{Pages 85--103}.
\newblock Boston, MA~(1972). Springer US.

\bibitem{clark1990unit}
Brent~N. Clark, Charles~J. Colbourn, and David~S. Johnson.
\newblock ``Unit disk graphs''.
\newblock \href{https://dx.doi.org/https://doi.org/10.1016/0012-365X(90)90358-O}{Discrete Mathematics {\bf 86}, 165--177}~(1990).

\bibitem{grotti2025practicalusecasesneutral}
Matteo Grotti, Sara Marzella, Gabriella Bettonte, Daniele Ottaviani, and Elisa Ercolessi.
\newblock ``Practical use cases of neutral atoms quantum computers''~(2025).
\newblock  \href{http://arxiv.org/abs/2510.18732}{arXiv:2510.18732}.

\bibitem{graham2022rydberggraph}
T.~M. Graham, Y.~Song, J.~Scott, C.~Poole, L.~Phuttitarn, K.~Jooya, P.~Eichler, X.~Jiang, A.~Marra, B.~Grinkemeyer, M.~Kwon, M.~Ebert, J.~Cherek, M.~T. Lichtman, M.~Gillette, J.~Gilbert, D.~Bowman, T.~Ballance, C.~Campbell, E.~D. Dahl, O.~Crawford, N.~S. Blunt, B.~Rogers, T.~Noel, and M.~Saffman.
\newblock ``Multi-qubit entanglement and algorithms on a neutral-atom quantum computer''.
\newblock \href{https://dx.doi.org/10.1038/s41586-022-04603-6}{Nature {\bf 604}, 457–462}~(2022).

\bibitem{dasilva2023column}
Wesley da~Silva~Coelho, Loïc Henriet, and Louis-Paul Henry.
\newblock ``Quantum pricing-based column-generation framework for hard combinatorial problems''.
\newblock \href{https://dx.doi.org/10.1103/physreva.107.032426}{Physical Review A{\bf 107}}~(2023).

\bibitem{vercellino2023bbqmis}
Chiara Vercellino et~al.
\newblock ``Bbq-mis: A parallel quantum algorithm for graph coloring problems''.
\newblock In 2023 IEEE International Conference on Quantum Computing and Engineering (QCE).
\newblock \href{https://dx.doi.org/10.1109/QCE57702.2023.10198}{Pages 141--147}.
\newblock Bellevue, WA, USA~(2023).

\bibitem{vercellino2025hybrid}
Chiara Vercellino, M.~Yassine Naghmouchi, Wesley Coelho, Giacomo Vitali, Alberto Scionti, Paolo Viviani, Olivier Terzo, and Bartolomeo Montrucchio.
\newblock ``Hybrid quantum-classical branch-and-price method for the vertex coloring problem''~(2025).
\newblock  \href{http://arxiv.org/abs/2508.18887}{arXiv:2508.18887}.

\bibitem{perron2025cg}
Cédrick Perron, Yves Bérubé-Lauzière, and Victor Drouin-Touchette.
\newblock ``Leveraging analog neutral atom quantum computers for diversified pricing in hybrid column generation frameworks''~(2025).
\newblock  \href{http://arxiv.org/abs/2510.04946}{arXiv:2510.04946}.

\bibitem{meng2023ddpp}
Zhiyi Meng, Yuting Zhou, Eldon~Y. Li, Xinying Peng, and Rui Qiu.
\newblock ``Environmental and economic impacts of drone-assisted truck delivery under the carbon market price''.
\newblock \href{https://dx.doi.org/https://doi.org/10.1016/j.jclepro.2023.136758}{Journal of Cleaner Production {\bf 401}, 136758}~(2023).

\bibitem{tarquini2024testing}
Sara Tarquini, Daniele Dragoni, Matteo Vandelli, and Francesco Tudisco.
\newblock ``Testing quantum and simulated annealers on the drone delivery packing problem''~(2024).
\newblock  \href{http://arxiv.org/abs/2406.08430}{arXiv:2406.08430}.

\bibitem{bettisorbelli2022greedy}
Francesco Betti~Sorbelli, Federico Cor\`{o}, Sajal~K. Das, Lorenzo Palazzetti, and Cristina~M. Pinotti.
\newblock ``Greedy algorithms for scheduling package delivery with multiple drones''.
\newblock In Proceedings of the 23rd International Conference on Distributed Computing and Networking.
\newblock \href{https://dx.doi.org/10.1145/3491003.3491028}{Page 31–39}.
\newblock ICDCN '22New York, NY, USA~(2022). Association for Computing Machinery.

\bibitem{10.1145/3571306.3571411}
Saswata Jana and Partha~Sarathi Mandal.
\newblock ``Approximation algorithms for drone delivery packing problem''.
\newblock In Proceedings of the 24th International Conference on Distributed Computing and Networking.
\newblock \href{https://dx.doi.org/10.1145/3571306.3571411}{Page 262–269}.
\newblock ICDCN '23New York, NY, USA~(2023). Association for Computing Machinery.

\bibitem{Silv_rio_2022}
Henrique Silvério, Sebastián Grijalva, Constantin Dalyac, Lucas Leclerc, Peter~J. Karalekas, Nathan Shammah, Mourad Beji, Louis-Paul Henry, and Loïc Henriet.
\newblock ``Pulser: An open-source package for the design of pulse sequences in programmable neutral-atom arrays''.
\newblock \href{https://dx.doi.org/10.22331/q-2022-01-24-629}{Quantum {\bf 6}, 629}~(2022).

\bibitem{gurobi}
{Gurobi Optimizer}.
\newblock ``Gurobi documentation''.
\newblock Online~(2020).
\newblock 2024-02-29.

\bibitem{hauke2020annealing}
Philipp Hauke, Helmut~G Katzgraber, Wolfgang Lechner, Hidetoshi Nishimori, and William~D Oliver.
\newblock ``Perspectives of quantum annealing: methods and implementations''.
\newblock \href{https://dx.doi.org/10.1088/1361-6633/ab85b8}{Reports on Progress in Physics {\bf 83}, 054401}~(2020).

\bibitem{doi:10.1287/ijoc.8.4.344}
Anuj Mehrotra and Michael~A. Trick.
\newblock ``A column generation approach for graph coloring''.
\newblock \href{https://dx.doi.org/10.1287/ijoc.8.4.344}{INFORMS Journal on Computing {\bf 8}, 344--354}~(1996).

\bibitem{HANSEN2009135}
P.~Hansen, M.~Labbé, and D.~Schindl.
\newblock ``Set covering and packing formulations of graph coloring: Algorithms and first polyhedral results''.
\newblock \href{https://dx.doi.org/https://doi.org/10.1016/j.disopt.2008.10.004}{Discrete Optimization {\bf 6}, 135--147}~(2009).

\bibitem{https://doi.org/10.1002/jgt.3190110403}
Zoltán Füredi.
\newblock ``The number of maximal independent sets in connected graphs''.
\newblock \href{https://dx.doi.org/https://doi.org/10.1002/jgt.3190110403}{Journal of Graph Theory {\bf 11}, 463--470}~(1987).

\bibitem{GRIGGS1988211}
Jerrold~R. Griggs, Charles~M. Grinstead, and David~R. Guichard.
\newblock ``The number of maximal independent sets in a connected graph''.
\newblock \href{https://dx.doi.org/https://doi.org/10.1016/0012-365X(88)90114-8}{Discrete Mathematics {\bf 68}, 211--220}~(1988).

\bibitem{Albash_2018}
Tameem Albash and Daniel~A. Lidar.
\newblock ``Adiabatic quantum computation''.
\newblock \href{https://dx.doi.org/10.1103/revmodphys.90.015002}{Reviews of Modern Physics{\bf 90}}~(2018).

\bibitem{openstreetmap}
{OpenStreetMap contributors}.
\newblock ``Planet dump retrieved from {OpenStreetMap} ({www.openstreetmap.org})''~(2025).
\newblock Accessed: 2025-07-08.

\bibitem{BREU19983}
Heinz Breu and David~G. Kirkpatrick.
\newblock ``Unit disk graph recognition is np-hard''.
\newblock \href{https://dx.doi.org/https://doi.org/10.1016/S0925-7721(97)00014-X}{Computational Geometry {\bf 9}, 3--24}~(1998).

\bibitem{vercellino2022neural}
Chiara Vercellino, Paolo Viviani, Giacomo Vitali, Alberto Scionti, Andrea Scarabosio, Olivier Terzo, Edoardo Giusto, and Bartolomeo Montrucchio.
\newblock ``Neural-powered unit disk graph embedding: qubits connectivity for some qubo problems''.
\newblock In 2022 IEEE International Conference on Quantum Computing and Engineering (QCE).
\newblock \href{https://dx.doi.org/10.1109/QCE53715.2022.00038}{Pages 186--196}.
\newblock ~(2022).

\bibitem{de_Correc_2025}
Christian de~Correc, Thomas Ayral, and Corentin Bertrand.
\newblock ``Approximate combinatorial optimization with rydberg atoms: The barrier of interpretability''.
\newblock \href{https://dx.doi.org/10.1103/ss3t-j4y5}{Physical Review A{\bf 112}}~(2025).

\bibitem{https://doi.org/10.1002/spe.4380211102}
Thomas M.~J. Fruchterman and Edward~M. Reingold.
\newblock ``Graph drawing by force-directed placement''.
\newblock \href{https://dx.doi.org/https://doi.org/10.1002/spe.4380211102}{Software: Practice and Experience {\bf 21}, 1129--1164}~(1991).

\bibitem{graph-draw}
Giuseppe~Di Battista, Peter Eades, Roberto Tamassia, and Ioannis~G Tollis.
\newblock ``Algorithms for drawing graphs: an annotated bibliography''.
\newblock \href{https://dx.doi.org/https://doi.org/10.1016/0925-7721(94)00014-X}{Computational Geometry {\bf 4}, 235--282}~(1994).

\bibitem{manetsch2024tweezerarray6100highly}
Hannah~J. Manetsch, Gyohei Nomura, Elie Bataille, Xudong Lv, Kon~H. Leung, and Manuel Endres.
\newblock ``A tweezer array with 6,100 highly coherent atomic qubits''~(2025).

\bibitem{SciPyProceedings11}
Aric Hagberg, Pieter Swart, and Daniel Chult.
\newblock ``Exploring network structure, dynamics, and function using networkx''.
\newblock \href{https://dx.doi.org/10.25080/TCWV9851}{Proceedings of the 7th Python in Science Conference}~(2008).

\bibitem{PhysRevApplied.23.064023}
S\'ebastien Perseguers.
\newblock ``Hardness-dependent quantum adiabatic schedules for the maximum-independent-set problem''.
\newblock \href{https://dx.doi.org/10.1103/PhysRevApplied.23.064023}{Phys. Rev. Appl. {\bf 23}, 064023}~(2025).

\bibitem{10.1108/IJLM-04-2023-0149}
Amer Jazairy, Emil Persson, Mazen Brho, Robin von Haartman, and Per Hilletofth.
\newblock ``Drones in last-mile delivery: a systematic literature review from a logistics management perspective''.
\newblock \href{https://dx.doi.org/10.1108/IJLM-04-2023-0149}{The International Journal of Logistics Management {\bf 36}, 1--62}~(2024).
\newblock  \href{http://arxiv.org/abs/https://www.emerald.com/ijlm/article-pdf/36/7/1/10056587/ijlm-04-2023-0149en.pdf}{arXiv:https://www.emerald.com/ijlm/article-pdf/36/7/1/10056587/ijlm-04-2023-0149en.pdf}.

\bibitem{app12010207}
Esrat~F. Dulia, Mir~S. Sabuj, and Syed A.~M. Shihab.
\newblock ``Benefits of advanced air mobility for society and environment: A case study of ohio''.
\newblock \href{https://dx.doi.org/10.3390/app12010207}{Applied Sciences{\bf 12}}~(2022).

\bibitem{johansson2012qutip}
J.R. Johansson, P.D. Nation, and Franco Nori.
\newblock ``Qutip: An open-source python framework for the dynamics of open quantum systems''.
\newblock \href{https://dx.doi.org/https://doi.org/10.1016/j.cpc.2012.02.021}{Computer Physics Communications {\bf 183}, 1760--1772}~(2012).

\bibitem{johansson2013qutip}
J.R. Johansson, P.D. Nation, and Franco Nori.
\newblock ``Qutip 2: A python framework for the dynamics of open quantum systems''.
\newblock \href{https://dx.doi.org/https://doi.org/10.1016/j.cpc.2012.11.019}{Computer Physics Communications {\bf 184}, 1234--1240}~(2013).

\bibitem{bidzhiev2025efficientemulationneutralatom}
Kemal Bidzhiev, Stefano Grava, Pablo~le Henaff, Mauro Mendizabal, Elie Merhej, and Anton Quelle.
\newblock ``{Efficient Emulation of Neutral Atom Quantum Hardware}''~(2025).
\newblock  \href{http://arxiv.org/abs/2510.09813}{arXiv:2510.09813}.

\bibitem{GOLUMBIC1980171}
Martin~Charles Golumbic.
\newblock ``Chapter 8 - interval graphs''.
\newblock In Martin~Charles Golumbic, editor, Algorithmic Graph Theory and Perfect Graphs.
\newblock \href{https://dx.doi.org/https://doi.org/10.1016/B978-0-12-289260-8.50015-7}{Pages 171--202}.
\newblock Academic Press~(1980).

\end{thebibliography}

\end{document}